\documentstyle[prl,aps,preprint,eqsecnum]{revtex}
\draft
\tighten

\begin{document}

\title{Second-quantization of open systems using quasinormal modes}

\author{K. C. Ho, P. T. Leung, Alec Maassen van den Brink and K. Young}

\address{Department of Physics, The Chinese University of Hong Kong,
Hong Kong, China}

\date{\today}

\maketitle

\begin{abstract}

The second-quantization of a scalar field in an open cavity is formulated,
from first principles, in terms of the quasinormal modes (QNMs),
which are the eigensolutions of 
the evolution equation that 
decay exponentially in time as energy leaks to the outside. 
This formulation provides a description involving 
the cavity degrees of freedom only, 
with the outside acting as a (thermal or driven) source. 
Thermal correlation functions and cavity Feynman propagators
are thus expressed in terms of the QNMs,
labeled by a discrete index rather than a continuous
momentum.  Single-resonance domination
of the density of states and the spontaneous decay rate is
then given a proper foundation.
This is a first essential step towards the 
application of QNMs to cavity QED phenomena, 
to be reported elsewhere.

\end{abstract}


\section{Introduction}
\label{intro}

In this paper we are concerned with quantum fields
in {\em open\/} cavities ---
the obvious example of ultimate interest would be
electromagnetic fields in optical cavities \cite{optcav}
and the associated problem of cavity quantum electrodynamics
(CQED).  Such systems are open because energy leaks
to the outside (e.g., via output coupling), and
as dissipative systems cannot normally be quantized
on their own \cite{diss}; rather, one must also consider the bath into which
energy escapes, so that the total universe is conservative.
Thus such cavities, say of linear dimension $a$, 
can be embedded in a universe of dimension $\Lambda \rightarrow \infty$.
One can then quantize on the modes of the universe,
which are labeled by a wavenumber $p$ spaced by
$\Delta p \sim \pi / \Lambda \rightarrow 0$.  The field quanta
are then created or destroyed by operators $a^{\dagger}(p)$ and
$a(p)$, and higher-order processes involve integrals $\int\! dp\, \cdots\;$.

On the other hand, these cavities often have a very small amount of
leakage, characterized by a parameter  
$\epsilon = Q^{-1} \ll 1$, where the quality factor of the cavity
can be as high as $Q \sim 10^6$ or more.  If this is the case,
the intuition developed from a {\em closed} cavity, also of length $a$, should be
relevant.  A closed cavity is a conservative system, with normal modes (NMs)
labeled by a {\em discrete} index $j = 1, 2, \cdots$, where the wavenumber  is
$p_j \sim j \pi / a$, $\Delta p \sim \pi / a$.  
Field quanta in such a closed cavity are created and
destroyed by operators $a^{\dagger}_j$ and $a_j$, and higher-order processes
involve discrete sums $\sum_j \cdots$.  Can quantum fields in an {\em open} cavity
be described in a similar way --- in terms of discrete modes and the corresponding
operators?  If this is possible, computations will be
simplified and will correspond to physical intuition, with
each term $j$ associated with a cavity ``mode".
The connection with
the limit of a closed cavity ($\epsilon \rightarrow 0$) would also become
manifest.

Quantization of a closed system relies on its NMs; the counterparts
in an open system are the quasinormal modes (QNMs), which are again
factorized solutions

\begin{equation}
\phi(x,t) = f_j(x)\mskip0.5\thinmuskip e^{-i\omega_jt}\;,
\label{eigensol}
\end{equation}

\noindent
where $\mathop{\rm Im}\nolimits \omega_j < 0$ because of the loss of energy. 
Each QNM corresponds to a resonance, with a width
$\gamma_j = | \mathop{\rm Im}\nolimits \omega_j|$.
The purpose of this paper, in short, is to develop a formalism whereby
field quantization can be implemented in terms of these QNMs,
and to define and study operators $a^{\dagger}_j$,
$a^{\vphantom{\dagger}}_j$ for these modes.  
Specifically, one wishes to express field correlation functions,
Feynman propagators and other quantities
in terms of the QNMs.  The dissipative nature of the system is
then contained in the QNMs
themselves.

The advantages for CQED would be obvious.  The simplest phenomenon to
which such a formalism can apply is the well-known enhancement (or suppression)
of spontaneous decay rates when the emitted radiation falls on (or between)
resonances \cite{cavQED}.
Resonance domination 
of these processes has been 
discussed starting with the heuristic argument due to Purcell~\cite{purcell}.
He proposed that the Fermi golden rule should be generalized:
the density of states per unit volume,
instead of the usual $d_0(\omega) = \omega^2/(\pi^2c^3)$ for vacuum
(where $c$ is the velocity of light),
is to be replaced by $d(\omega) \sim D/(2\gamma V)$
for a $D$-fold degenerate QNM of width $\gamma$ in a cavity of
volume $V$. This leads to an 
enhancement factor of $K = d/d_0 \sim (1/8\pi)DQ (\lambda^3/V)$ 
for spontaneous emission on resonance, where $\lambda$ 
is the wavelength of light emitted
and $Q$ is the quality factor of the cavity. 
The essence of this argument is that each resonance counts as
one state, i.e., in a suitable sense it carries unit weight.
While intuitively plausible, this statement is
difficult to justify formally --- since
the entire concept of a state, i.e., an NM, falls apart in an
open system.
However, this argument, and its many variants and extensions, would find
natural expression in a framework that quantizes on the QNMs,
and we shall in particular show explicitly below that each resonance
carries unit weight.

In Section \ref{class}, the QNM expansion of {\em classical}\/
fields outgoing from a cavity is reviewed.  
The classical results may be organized into two levels.
First, under conditions to be specified,
the Green's function $G$ can be expanded in terms
of QNMs.  Second, one can try to expand
the outgoing classical fields $\phi$ in terms of QNMs, and to establish a
linear space structure similar to that for conservative systems.
In order to do so, it turns out to be necessary to make use of
a two-component formalism, dealing with $\phi$ and the
conjugate momentum ${\hat \phi}$ at the same time.  The 
linear space formalism is more elegant, but
in its simplest form is limited to only 1 d.

The quantum formalism can likewise be approached in two ways.
In the first, which we shall call the Green's function
method (Section \ref{1comp}), one focuses on $c$-number 
correlation functions and propagators
without explicitly expanding the
field operator $\phi$ in terms of QNMs.  
The key idea is that the retarded propagator $G^{\rm R}$
of the quantum theory is exactly the same as the classical Green's function
$G$, and the latter has a QNM expansion.  Once $G^{\rm R}$ is obtained,
it is straightforward to derive a QNM expansion for the correlation function
$F$ as well.

In the second, which we shall call the field expansion method,
one tries to establish an expansion of the quantum
field $\phi$, in parallel with the linear space structure established for
classical fields.  This allows us to interpret the 
expansion coefficients $a^{\dagger}_j$ and $a_j$,
roughly speaking, as generalized creation 
and annihilation operators for the discrete QNMs $j$.
However, before doing so, it has to be recognized that quantum fields cannot
be constrained by the outgoing wave condition --- for the
simple reason that zero-point (and thermal) fluctuations must contain an
incoming component.
Thus, the first step in developing this method,
presented in Section \ref{incoming},
is to generalize the 
field expansion to handle {\em incoming} waves as well.
With this generalization, one can then subject
the fields to canonical quantization
in Section \ref{quant}.
This is done by starting with the universe, 
a closed hermitian system for 
which the quantization is unambiguously defined. 
Then, in parallel with the usual removal 
of bath oscillators \cite{diss},
the outside degrees of freedom are 
eliminated from the equations of motion~\cite{fn3}. 
The results will be equations of motion and commutation
relations for the {\em discrete}\/ operators
$a^{\dagger}_j$ and $a_j$, in which the effects of the outside
bath are clearly displayed: the loss of energy of each mode
by leakage, and the pumping of each mode by the thermal
or quantum fluctuations from the outside.

The formalism is then used
to evaluate the correlation function $F$ in Section \ref{correl},
and the results are compared with those obtained from the
Green's function method.  Interestingly, the results appear to be different ---
those derived from the Green's function method contain a single sum $\sum_j$
over the QNMs, while the field expansion method yields a double sum
$\sum_{jk}$ with off-diagonal terms.  The two are, however, shown to
be equal through an identity on $G^{\rm R}$.  
Recalling that the expansion of the classical field is unique 
only when the second component~${\hat \phi}$ is considered
at the same time, we next show that the expansion of the correlation
function $F$ is also unique if we consider ${\hat \phi}$ as well, 
giving the non-diagonal form.
The density of states~$d$, which is intimately related to the
correlation function, is also expressed in terms of QNMs; in particular
it is shown that up to corrections of $O(Q^{-1})$, each resonance
carries unit weight in the density of states.

The results on the correlation function are then used, in Section \ref{feyn}, 
to evaluate and discuss the Feynman propagator $G^{\rm F}$, which is the
fundamental building block for CQED.  Again, equivalent
diagonal and non-diagonal forms are obtained.
Particular attention is paid to the
equal space propagator ${\tilde G}^{\rm F}(x,x,\omega)$, 
whose imaginary part is related to the life-time
of an excited atom placed at $x$.  This quantity is discussed in the
approximation of
domination by a single resonance, providing justification for 
Purcell's heuristic argument \cite{purcell} on
the enhancement of spontaneous decay rates.  The advantage of using
the non-diagonal expression is again emphasized.

In Section \ref{example}, a very simple example is studied
explicitly, and its correlation function and energy density are
expressed in terms of
a sum over QNM contributions.

Some final remarks are then
given in Section \ref{remarks}.  We stress that this paper is
concerned entirely with {\em free} fields, either as a model
of the free electromagnetic field in an optical cavity, or as
the zeroth-order building blocks in an interacting theory, e.g.,
the propagators as ingredients in higher-order Feynman diagrams.
The development of the interacting theory and its application
to CQED phenomena will be given elsewhere \cite{partII}.
A partial account of the present theory has been given in \cite{KL}.

\section{Classical fields}
\label{class}

In this Section, we summarize the QNM expansion
for classical fields.  In this paper, we deal with 
scalar fields in 1 d only.

For closed, linear systems,
eigenfunction expansions, based on 
the eigenfunctions or NMs of 
their evolution operators,
are a tool of 
vital importance in theoretical physics. 
However, open systems are 
not directly amenable to an NM analysis.
Examples of open systems include optical cavities \cite{optcav},
and finite regions of space near
astrophysical objects, from which gravitational waves
can escape \cite{grav,kg}. 
In these systems, any initial state decays in time,
so stationary NMs do not exist. 
As the simplest example, we shall be concerned with 
the scalar wave equation in one space dimension,

\begin{equation}
\rho(x)\mskip0.5\thinmuskip\partial_t^2\phi = \partial_x^2\phi
\label{wave-eq}
\end{equation}

\noindent
studied in a ``cavity" $0\le x\le a$, with the nodal 
boundary condition

\begin{equation}
\phi(x{=}0,t)=\dot{\phi}(x{=}0,t) = 0
\label{bc-left}
\end{equation}

\noindent
at one end but with the outgoing one

\begin{equation}
\dot{\phi}(a^+,t) = -\phi'(a^+,t)
\label{bc-right}
\end{equation}

\noindent
at the other.   The latter condition states that, 
just outside the cavity boundary, 
the field $\phi(x,t)$ is an outgoing wave $\phi(x-t)$; 
the condition is specified at $a^+$ because, as we shall see 
below, one is often concerned with models in which there is a 
singularity in $\rho(x)$ at $x=a$, 
leading to possible discontinuities in $\phi(x)$ or $\phi'(x)$.
The boundary condition (\ref{bc-right}) turns the cavity into a dissipative system 
that is leaky but not absorptive.
The model (\ref{wave-eq}) has been widely used as the scalar model
of electromagnetism in an optical cavity \cite{optcav}.
More physically, the 1-d nature is realized in Fabry-Perot cavities
with lengths much smaller than the lateral dimensions, and
the scalar field model is rigorously applicable to the transverse electric
sector.

For the system (\ref{wave-eq})--(\ref{bc-right}), 
the eigensolutions, labeled by an index $j$, have the form
(\ref{eigensol}),
with the QNMs or cavity resonances $f_j$ satisfying

\begin{equation}
[\partial_x^2+\rho(x)\mskip0.5\thinmuskip\omega_j^2]
\mskip0.5\thinmuskip f_j = 0
\label{def-QNM}
\end{equation}

\noindent
and the boundary conditions (\ref{bc-left}), 
(\ref{bc-right}) translating to

\begin{equation}
f_j(0) = 0,  \qquad   f_j'(a^+) = i\omega_jf_j(a^+)\;.
\label{f-bc}
\end{equation}

\noindent
It is easily verified that $\mathop{\rm Im}\nolimits \omega_j <0$,
so that the solution 
(\ref{eigensol}) is indeed decaying in time. 
Furthermore, the frequencies $\omega_j$, which we suppose to be 
ordered according to increasing real parts, are spaced 
by $\Delta\omega\sim\pi/a$, approximately as for a conservative
system of size $a$. 
With the possible exception of
modes with $\mathop{\rm Re}\nolimits \omega_j =0$,
the QNMs always occur in pairs 
with $\omega_{-j}^{\vphantom{*}}=-\omega_j^*$, and 
one can choose $f_{-j}^{\vphantom{*}}=f_j^*$.

The usual formalism concerning eigenfunction 
expansions relies on the hermiticity of the evolution operator, 
which only holds in the conservative case, and therefore 
breaks down for open systems. One possible resolution 
is to embed the cavity into a ``universe"
$0\le x\le\Lambda$ with a nodal condition
at $x=\Lambda\rightarrow\infty$, and 
study its NMs --- the modes of the universe. Namely, 
the system (\ref{wave-eq})--(\ref{bc-right}) is the 
restriction to $x\le a$ of the problem (\ref{wave-eq}) 
on the half line $0\le x<\infty$, if one sets

\begin{equation}
\rho(x{>}a) \equiv 1
\label{rho-out}
\end{equation}

\noindent
and with the extension of 
the initial conditions to the ``outside" $x>a$ obeying 
$\phi'(x{>}a,t{=}0)=-\dot{\phi}(x{>}a,t{=}0)$. 
However, this has the obvious disadvantage of having 
to work with a continuum of states
(spaced by $\Delta \omega \sim \pi / \Lambda \rightarrow 0$)
as opposed to the 
discrete set of eigenfunctions in the conservative case. 
Besides, the closed system of equations
(\ref{wave-eq})--(\ref{bc-right}) shows that 
even in the presence of dissipation the time 
evolution of the cavity can be studied {\em without}\/ 
explicit reference to the outside, which is the principal
goal of the program of second quantization of the open system.

Previous work (see \cite{lly,LTY1,LTY2} and references therein) 
has established that, in spite of the lack of hermiticity 
in the conventional sense, an eigenfunction expansion for outgoing waves
in classical open wave systems can be formulated in terms of the cavity degrees of 
freedom only, overcoming the disadvantages 
of the modes of the universe approach. The sufficient conditions for 
this QNM expansion are as follows.

\begin{itemize}
\item[(a)]The function $\rho(x)$ has at least a step discontinuity at $x=a$.
This demarcates a well-defined cavity region.
\item[(b)]The function $\rho(x)$ has no tail outside the cavity,
i.e., $\rho(x{>}a)\equiv1$. This 
condition ensures that the outside does not reflect
outgoing waves back into the cavity, enabling the 
complete elimination of the environment from the equations of motion.
\end{itemize}

\noindent
These conditions are satisfied for optical cavities bounded from
extended vacuum by a sharp material interface.

The completeness of the QNMs can be pursued at two levels.
First, one shows that the retarded Green's 
function of the system has the representation

\begin{equation}
G(x,y,t) =
\sum_j \frac{f_j(x)\mskip0.5\thinmuskip f_j(y)}{2i\omega_j}
\mskip0.5\thinmuskip  e^{-i\omega_jt}
\label{greenfie}
\end{equation}

\noindent
for $0 \le x,y \le a$ and $t \ge 0$, where the $f_j$'s are 
normalized according to (\ref{normal}) below. 
Thus, the dynamics is contained entirely in the QNMs, 
leading to a simple method of obtaining
the retarded propagators and quantum correlation functions,
as sketched in Section \ref{1comp}.

Second, realizing that the wave equation (\ref{wave-eq}),
like any classical hamiltonian problem, requires both 
position and momentum to be specified as initial data, 
one introduces function {\em pairs}\/ 
$\bbox{\phi}=(\phi,\hat{\phi})^{\rm T}$ with 
the conjugate momentum $\hat{\phi}\equiv\rho\dot{\phi}$, 
so that for eigenfunctions 
$\bbox{f}_{\mskip-0.5\thinmuskip\!j}=(f_j,-i\rho\omega_jf_j)^{\rm T}$. 
The space of all function pairs satisfying 
the boundary conditions (\ref{bc-left}) 
and~(\ref{bc-right}) will be denoted as $\Gamma$ ---
the space of outgoing waves.

Using these pairs, one can prove that the time evolution 
generated by (\ref{greenfie}) can be 
recast in the form \cite{fn1} 

\begin{equation}
\bbox{\phi}(t) = \sum_j a_j(t) \bbox{f}_{\mskip-0.5\thinmuskip\!j}\;,
\label{evolve}
\end{equation}

\noindent
where the expansion coefficients are given by

\begin{equation}
 a_j(t) = \frac{1}{2\omega_j}
 \langle \bbox{f}_{\mskip-0.5\thinmuskip\!j},\bbox{\phi}(t) \rangle
\label{a_n1}
\end{equation}

\noindent
with $a_j(t)=a_j(0)\mskip0.5\thinmuskip e^{-i\omega_jt}$ and the
{\em bilinear scalar product}\/

\begin{equation}
\langle\bbox{\phi},\bbox{\chi}\rangle = i \left\{
  \int_0^{a^+}\!\!dx\,\left[ \phi(x)\mskip0.5\thinmuskip\hat{\chi}(x)
  +\hat{\phi}(x)\mskip0.5\thinmuskip\chi(x)\right]
  +\phi(a^+)\mskip0.5\thinmuskip\chi(a^+) \right\} \;.
\label{scalarprod}
\end{equation}

\noindent
By simply letting $t\!\downarrow\!0$ in 
(\ref{evolve}) one arrives at a 
{\em two-component expansion}\/ for an 
arbitrary $\bbox{\phi}\in\Gamma$ \cite{fn2,messiah}.
This expansion makes the completeness of the QNMs manifest.
The normalization used in (\ref{greenfie}) to (\ref{a_n1}) 
can be concisely expressed as

\begin{equation}
\langle\bbox{f}_{\mskip-0.5\thinmuskip\!j},
\bbox{f}_{\mskip-0.5\thinmuskip\!j}\rangle=2\omega_j\;.
\label{normal}
\end{equation}

\noindent
It is seen that (\ref{normal}) in general 
is not real, underlining the difference between 
the product (\ref{scalarprod}) and a conventional 
one involving complex conjugation.
The fact that (\ref{normal}) is bilinear also
serves to establish a phase convention for
the wavefunctions.

Upon introducing the two-component evolution operator

\begin{equation}
\bbox{H}=i \pmatrix{ 0 & \rho(x)^{-1} \cr \partial_x^2 & 0}
\;,
\label{harray}
\end{equation}

\noindent
the cavity evolution (\ref{wave-eq}) can be written 
as $i\mskip0.5\thinmuskip\partial_t\bbox{\phi}=\bbox{H}\bbox{\phi}$, 
in striking analogy with quantum mechanics. 
In this notation, the definition (\ref{def-QNM}) 
of $f_j$ takes the form $\bbox{H}\bbox{f}_{\mskip-0.5\thinmuskip\!j}
=\omega_j\bbox{f}_{\mskip-0.5\thinmuskip\!j}$. 
The operator $\bbox{H}$ can be shown to be symmetric 
with respect to the form (\ref{scalarprod}), i.e.,

\begin{equation}
\langle \bbox{\phi},\bbox{H}\bbox{\chi} \rangle =
\langle \bbox{\chi},\bbox{H}\bbox{\phi} \rangle
\end{equation}

\noindent
for any $\bbox{\phi},\bbox{\chi} \in \Gamma$. 
This analog of hermiticity
holds even though the system is not conservative.
The symmetry of $\bbox{H}$ yields the orthogonality relation

\begin{equation}
\langle\bbox{f}_{\mskip-0.5\thinmuskip\!j},
\bbox{f}_{\mskip-0.5\thinmuskip\!k}\rangle = 0
\qquad   \text{for $\omega_j \neq \omega_k$}
\label{ortho}
\end{equation}

\noindent
in an immediate transcription of the usual proof,
leading to the uniqueness of the expansion.
Incidentally, an expansion such as (\ref{evolve})
but involving the first component alone would not be unique.

It is appropriate to contrast the two methods of approaching the classical theory,
since they respectively underpin the two methods for dealing with
the quantum case. 
The expansion of the Green's function is easy to derive,
and is readily generalized to higher dimensions; however,
in itself it does not lead to a {\em unique}\/ expansion of
the field, nor to concepts of orthogonality.  The two-component
approach based on (\ref{evolve}) is more elegant,
exhibits a deeper resemblance to conservative systems,
and most importantly leads to a {\em unique}\/ expansion
in terms of orthogonal functions.  The two-component
expansion can in principle be generalized to higher dimensions
by treating each angular momentum sector $l$ as a 1-d radial
problem~\cite{out}, but the degree of complexity increases with $l$.
Thus each method has its own merits; both will be pursued below,
and the results compared.

\section{Green's function method}
\label{1comp}

The quantum mechanics of the system is specified by the hamiltonian

\begin{equation}
H= \int_0^{\infty} \!\! dx\, h(x) = \int_0^{\infty}\!\!dx\,
    \left[ \frac{{\hat{\phi}\vphantom{\overline{\phi}}}^2}{2\rho}
    +\frac{1}{2} \left(\frac{\partial  \phi}{\partial x} \right)^{\!\!2} \right]
\label{def-H}
\end{equation}

\noindent
together with the 
canonical equal-time commutation relation 

\begin{equation}
[\phi(x),\hat{\phi}(y)]=i\delta(x-y) \;.
\label{ccr}
\end{equation}

\noindent
Time evolution is then generated by
means of the Heisenberg equation $\dot{A}=i[H,A]$ 
for an arbitrary operator $A$.
However, instead of the equations of motion for the quantum operators,
in this Section we focus first on
the retarded propagator

\begin{equation}
G^{\rm R}(x,y,t) = -i\theta(t)\langle[\phi(x,t),\phi(y)]\rangle \;,
\label{def-gr}
\end{equation}

\noindent
in which $\phi$ is of course to be regarded as an operator,
and $\langle \cdots \rangle$ denotes the expectation value
at a finite temperature $T = 1/ \beta$;
throughout we take $\hbar=k_{\rm B}=1$.

The central idea is that this propagator defined in terms
of the quantum fields can be evaluated
without explicitly introducing an expansion for the field operators $\phi$,
by simply noticing that $G^{\rm R}(x,y,t)$ is exactly the same as the
Green's function $G$ of the classical wave equation \cite{AGD},
which has the expression (\ref{greenfie}) 
in terms of QNMs.   This relationship between $G^{\rm R}$ and $G$
follows from the commutation relation (\ref{ccr}).

In terms of $G^{\rm R}$, it is straightforward to compute
the equilibrium correlation function

\begin{equation}
F(x,y,t)\equiv\langle\phi(x,t)\mskip0.5\thinmuskip\phi(y)\rangle\;.
\label{def-F}
\end{equation}

\noindent
We shall devote attention to $F$,
because the physical quantities of interest in quantum field theory
can often be formulated in terms of 
correlation functions, at either zero or finite temperatures. 
For example, the Casimir force is merely the vacuum 
expectation value of the electromagnetic stress tensor, which is 
an equal-time equal-space correlation function of two field operators.
The spontaneous decay rate of an atom in an excited state is, in the golden
rule approximation, related to the correlation
function of two electric field operators.

Since the correlation function is related to the retarded propagator \cite{AGD},
one gets

\begin{eqnarray}
\tilde{F}(x,y,\omega) &=& \frac{-2}{1-e^{-\beta\omega}}
\mathop{\rm Im}\nolimits\tilde{G}^{\rm R}(x,y,\omega)
\label{relfg} \\
&=& 
\frac{i\omega}{1-e^{-\beta\omega}}
\sum_j \frac{f_j(x)f_j(y)}
    {\omega_j(\omega_{\vphantom{j}}^2-\omega_j^2)} \;.
\label{F-d}
\end{eqnarray}

\noindent
The real-time correlator
can be obtained from (\ref{F-d}) by contour 
integration, yielding

\begin{eqnarray}
  F(x,y,t)&=&\sum_j\frac{f_j(x)f_j(y)}{2\omega_j(1-e^{-\beta\omega_j})}
    e^{-\beta\omega_j\theta(-t)-i\omega_j|t|}\nonumber\\
  &&+\sum_{m=1}^\infty\frac{e^{-\mu_m|t|}}{\beta}
  \left[\tilde{G}^{\rm R}(x,y,-i\mu_m)
        -\tilde{G}^{\rm R}(x,y,i\mu_m)\right]\;.
\label{F-realt}
\end{eqnarray}

\noindent
The first term in this formula is due to the 
QNM poles in $\tilde{F}(x,y,\omega)$; the second term, 
which has no counterpart in $G^{\rm R}(x,y,t)$, is caused 
by the Matsubara poles in $\tilde{F}(x,y,\omega)$ at 
frequencies $\mu_m\equiv2\pi mT$.

This very simple derivation has the advantage that it goes through in 
situations where the two-component formalism 
may be more complicated.

In principle, physical quantities can be expressed in terms of $F$ ---
bilinear quantities (such as the energy density) as linear combinations
of $F$ and its derivatives, and other quantities involving products of $F$'s.
For example, the energy density is

\begin{equation}
\langle h(x) \rangle =
  \frac{1}{2} \left[ -\rho(x) \partial_t^2 + \partial_x \partial_y \right]
\, F(x,y,t) \Big|_{x=y,t=0}\;.
\label{eden1}
\end{equation}

\noindent
However, this quantity is divergent.
Subtracting off the zero-point, we  consider

\begin{eqnarray}
U(x,T) &=& \langle h(x) \rangle - \langle h(x) \rangle_{T=0}
\nonumber \\
&=& \frac{1}{2} \left[ -\rho(x) \partial_t^2 + \partial_x \partial_y \right]
\, F_{\rm S}(x,y,t) \Big|_{x=y,t=0}
\label{edeb2}
\end{eqnarray}

\noindent
in terms of the subtracted correlation function $F_{\rm S} \equiv F - F_0$,
where $F_0 = \lim_{\beta \rightarrow \infty} F$.
The limit $\beta \rightarrow \infty$ is best taken
in (\ref{F-d}) prior to Fourier inversion.
We further make use of the expansion for ${\tilde G}^{\rm R}$
to get

\begin{eqnarray}
F_{\rm S}(x,y,t)&=&
\sum_j\frac{f_j(x)f_j(y)}{\omega_j} C_j(t) \;, \\
C_j(t)&=& \frac{e^{-i\omega_j t}}{2}
\left[\frac{1}{e^{\beta\omega_j}-1}
 + \theta(-\mathop{\rm Re}\nolimits\omega_j)\right]
+ \frac{i}{\beta}
\sum_{m=1}^\infty \frac{\mu_m e^{-\mu_m t}} {\mu_m^2 + \omega_j^2}
\nonumber\\
&& -{}\frac{i}{4\pi}e^{i\omega_jt} {\rm E}_1(i\omega_jt) 
-{}\frac{i}{4\pi}e^{-i\omega_jt} {\rm E}_1(-i\omega_jt)
\;,
\label{FR2-realt}
\end{eqnarray}

\noindent
where this and subsequent formulas for $F_{\rm S}$ are written
only for $t>0$. ${\rm E}_1(z)$ is the exponential integral function~\cite{A&S}

\begin{equation}
{\rm E}_1(z) = \int_z^{+\infty} \frac{e^{-u}}{u}\, du\;,
\label{expfn}
\end{equation}

\noindent
in which the integration contour is defined not to pass through the
origin and the negative real axis; on that semi-axis, the function is
defined as the principal value. We further define
$\theta(0)\equiv\frac{1}{2}$.

Alternatively, for greater formal similarity to the conservative case,
(\ref{FR2-realt}) can be rewritten as
\begin{eqnarray}
F_{\rm S}(x,y,t)&=&
{\sum_{\!\!\!\mathop{\rm Re}\nolimits\omega_j\ge0}\!\!\!}'\;
2\mathop{\rm Re}\nolimits\left[\frac{f_j(x)f_j(y)}{\omega_j}
\tilde{C}_j(t)\right] \;,\label{FS-realt} \\
\tilde{C}_j(t)&=& \frac{e^{-i\omega_j t}}{2(e^{\beta\omega_j}-1)} 
+ \frac{i}{\beta}
\sum_{m=1}^\infty \frac{\mu_m e^{-\mu_m t}} {\mu_m^2 + \omega_j^2}
\nonumber\\
&& -{}\frac{i}{4\pi}e^{i\omega_jt} {\rm E}_1(i\omega_jt) 
-{}\frac{i}{4\pi}e^{-i\omega_jt} {\rm E}_1(-i\omega_jt)
\;.
\end{eqnarray}
The prime on the sum in (\ref{FS-realt}) signifies that terms with
$\mathop{\rm Re}\nolimits\omega_j=0$ are to be taken with weight $\frac{1}{2}$.

The actual evaluation of $U(x,T)$ needs
some care in the $j \rightarrow \infty$ part of the sums.
These details, and the very similar
problem for the calculation of the Casimir force, will be given elsewhere. 
\section{Incoming waves}
\label{incoming}

The expansion of a classical field sketched in
Section \ref{class} is restricted to outgoing waves, i.e., 
to $\bbox{\phi} \in \Gamma$, satisfying (\ref{bc-right}).
In preparing the ground for the expansion of a {\em quantum} field,
it is necessary to remove this restriction, for the
simple reason that the zero-point quantum fluctuations
will inevitably contain incoming waves as well.  Moreover, one
would wish that the ensuing theory should be applicable
to situations where there is an incoming pump field.

Thus, we study the wave equation (\ref{wave-eq})
for the system together with the outside ``bath", i.e., 
on the half line $x>0$, with $\rho(x)$ satisfying (\ref{rho-out}) and the 
boundary condition (\ref{bc-left}).
The initial 
conditions are now arbitrary and accordingly
the outgoing boundary condition (\ref{bc-right}) is abandoned,
i.e., the restriction of $\bbox{\phi}$ to the cavity 
need not lie in $\Gamma$. For the outside $x>a$ 
(where $\rho(x)=1$) the initial data are decomposed as

\begin{equation}
\bbox{\phi}(x{>}a,0) = \bbox{\phi}_{\rm IN} + \bbox{\phi}_{\rm OUT}\;,
\end{equation}

\noindent
with $\bbox{\phi}_{\rm IN}$ satisfying the incoming 
wave condition $\phi_{\rm IN}'=\hat{\phi}_{\rm IN}^{\vphantom{\prime}}$, 
while $\phi_{\rm OUT}'=-\hat{\phi}_{\rm OUT}^{\vphantom{\prime}}$. 
For the cavity subsystem this decomposition leads to 
the boundary condition

\begin{eqnarray}
\phi'(a^+\!,t) + \hat{\phi}(a^+\!,t)
& = & 2\hat{\phi}_{\rm IN}(a+t) \nonumber\\
& \equiv&b(t)\;,
\label{def-b}
\end{eqnarray}

\noindent
where the driving force $b$ (see (\ref{eqom-a_n}) 
below for its name), being determined by the initial data, 
is supposedly a known function (at least in a statistical sense)
which characterizes the waves incoming from the outside. 
Inside the cavity, the field is then expanded 
in terms of QNMs by (\ref{evolve}) with

\begin{eqnarray}
a_j &=& \frac{1}{2\omega_j}
 \left<\bbox{f}_{\mskip-0.5\thinmuskip\!j},\bbox{\phi}\right>
\nonumber\\
&=& \frac{i}{2\omega_j}
\left\{ \int_0^{a^+}\!\!dx\,f_j(x)
    \left[ \hat{\phi}(x)-i\rho(x)\mskip0.5\thinmuskip\omega_j\phi(x) \right]
         +f_j(a^+)\mskip0.5\thinmuskip\phi(a^+) \right\}\;.
\label{def-a_n}
\end{eqnarray}

\noindent
That is, we {\em retain}\/ the expansion formula and 
the inner product definition and notation even though 
$\bbox{\phi}\not\in\Gamma$.  As a consequence, 
the sum in (\ref{evolve}) will in general not 
converge to $\hat{\phi}$ at $x=a$, the point where the 
boundary condition is imposed.  Nevertheless, the sum for the first component
converges to $\phi$ everywhere, while the sum for the second component
converges to ${\hat \phi}$ everywhere except at $x=a$ \cite{fn4}.
(This is most easily appreciated by noticing that upon
changing ${\hat \phi}$ at just one point, the resultant
wavefunction can be made to lie in $\Gamma$.)
This flaw on a set 
of measure zero does not lead to problems, however, 
for the projection formula (\ref{def-a_n}) renders the coefficients
$a_j(t)$ well-defined irrespective of the convergence of 
the series (\ref{evolve}).

The equation of motion for $a_j$, which will survive
quantization, will now be derived.
By differentiating (\ref{def-a_n}) with respect to time, 
and then integrating by parts, one obtains 

\begin{equation}
\dot{a}_j(t)+i\omega_ja_j(t)
 = \frac{i}{2\omega_j}f_j(a^+)\mskip0.5\thinmuskip b(t)\;.
\label{eqom-a_n}
\end{equation}

\noindent
In contrast to the case of pure outgoing waves,
there is now an extra term on the right hand side:
each QNM is driven by
the ``force" $b(t)$, and at the same time decays because
of $\mathop{\rm Im}\nolimits \omega_j$.  The coupling to the ``force"
is determined by the surface value of the 
QNM wavefunction~$f_j(a^+)$.

\section{The field expansion method}
\label{quant}

Another approach to second-quantization
proceeds more explicitly by first promoting $\phi$ and ${\hat \phi}$
to operators \cite{fn5}.  These fields may be regarded
as operators for the entire ``universe", which is a conservative
system to which canonical quantization can be applied.
The same projection 
formula (\ref{def-a_n}) as in the classical case now 
defines the $a_j$'s as Hilbert space operators, obeying 
the equation of motion (\ref{eqom-a_n}).

The crucial point is that the field commutation relation
(\ref{ccr}) and the projection formula (\ref{def-a_n})
now lead directly to commutators for these coefficients, viz.,

\begin{eqnarray}
  [a_j,a_k]&=&\frac{1}{4\omega_j\omega_k}
  \left[\left<\bbox{f}_{\mskip-0.5\thinmuskip\!j},\bbox{\phi}\right>,
  \left<\bbox{f}_{\mskip-0.5\thinmuskip\!k},\bbox{\phi}\right>\right]\nonumber\\
  &=&-\frac{1}{4\omega_j\omega_k}
    \biggl\{\int_0^{a^+}\!\!dx\mskip0.5\thinmuskip dy\,
    \left(f_j(x)\mskip0.5\thinmuskip\hat{f}_k(y)[\hat{\phi}(x),\phi(y)]
    +\hat{f}_j(x)\mskip0.5\thinmuskip f_k(y)
    [\phi(x),\hat{\phi}(y)]\right)\nonumber\\
    &&\hphantom{-\frac{1}{4\omega_j\omega_k}\biggl\{}
    +\int_0^{a^+}\!\!dx\,f_j(x)\mskip0.5\thinmuskip
    f_k(a^+)[\hat{\phi}(x),\phi(a^+)]
     +\int_0^{a^+}\!\!dy\,f_j(a^+)\,f_k(y)[\phi(a^+),\hat{\phi}(y)]
    \biggr\}\;.\nonumber \\
&& \label{a_n-comm}
\end{eqnarray}

\noindent
In these equations, $\phi$ and ${\hat \phi}$ are
$q$-numbers, while $f_j$, $f_k$ are $c$-cumbers.
The two surface terms on the last line cancel as long 
as the delta function at the boundary of the integration 
interval is interpreted consistently.  In the first line,
the commutation relation (\ref{ccr})  
gives $\delta(x-y)$ and cancels one integration.
One is then left with

\begin{eqnarray}
  [a_j,a_k]&=&\frac{\omega_k-\omega_j}{4\omega_j\omega_k}
  \int_0^{a^+}\!\!dx\,\rho(x)\mskip0.5\thinmuskip
  f_j(x)\mskip0.5\thinmuskip f_k(x)
\label{comm-res}\\
  &=&\frac{i(\omega_j-\omega_k)f_j(a^+)\mskip0.5\thinmuskip f_k(a^+)}
       {4\omega_j\omega_k(\omega_j+\omega_k)}\;,
\label{comm-r2}
\end{eqnarray}

\noindent
where the second form follows from the first by means of 
the orthogonality relation (\ref{ortho}), and will be useful later  
for comparison with results from Section \ref{correl}.

The linear space structure for open systems involves
projections based on the generalized inner product (\ref{scalarprod})
which is bilinear rather than
linear in one vector and conjugate linear
in the other; thus the expression in (\ref{comm-res})
involves an integral over $ f_j(x)\mskip0.5\thinmuskip f_k(x)$ without
complex conjugation.  However, for the sake of a more
transparent comparison with the conservative case,
it is useful to re-write these expressions 
by changing $j \mapsto -j$ and using
$a_{-j} = a^{\dagger}_j$, $\omega_{-j} = -\omega^*_j$
and $f_{-j} = f^*_j$ to give:

\begin{eqnarray}
  [a^{\dagger}_j, a_k] &=&
-\frac{\omega_k+\omega^*_j}{4\omega^*_j \omega_k}
  \int_0^{a^+}\!\!dx\,\rho(x)\mskip0.5\thinmuskip
  f^*_j(x)\mskip0.5\thinmuskip f_k(x)
\label{comm-resa} \\
  &=&-\frac{i(\omega^*_j+\omega_k)f^*_j(a^+)\mskip0.5\thinmuskip f_k(a^+)}
       {4\omega^*_j\omega_k(\omega^*_j-\omega_k)} \;.
\label{comm-r2a}
\end{eqnarray}

\noindent
The result in the form (\ref{comm-resa}) reveals the conservative limit
most clearly; in this limit
the integral would simply be $\delta_{|j|,|k|}$. 

Comparison of  (\ref{comm-resa}) and (\ref{comm-r2a})
shows that 

\begin{equation}
R_j \equiv  \frac{ |f_j(a^+)|^2 } { 2 | \mathop{\rm Im}\nolimits \omega_j | }
 \rightarrow 1
\label{ratiolim}
\end{equation}

\noindent
in the conservative limit.
A more explicit proof is given in Appendix \ref{ident-app}.

The above commutators show that, if we define, for $j>0$ 

\begin{eqnarray}
\alpha_j &=& \sqrt{2\omega_j}a_j
\nonumber \\
\alpha_j^{\dagger}&=& \sqrt{2\omega_j^*}a_{-j} \;,
\label{alpha}
\end{eqnarray}

\noindent
then in the conservative limit these should reduce to the annihilation 
and creation operators, respectively \cite{fn6}. Indeed, the QNM expansion 
(\ref{evolve}) then takes the form

\begin{equation}
 \pmatrix{ \phi \cr \hat{\phi}}  = \sum_{j>0}
  \pmatrix{ (\alpha_j^{\dagger}+\alpha_j^{\vphantom{\dagger}})/
  \sqrt{2\omega_j} \cr
    i\rho\sqrt{\omega_j/2}(\alpha_j^{\dagger}-\alpha_j^{\vphantom{\dagger}}) } 
 \, f_j  \;,
\label{nm-exp}
\end{equation}

\noindent
the standard normal-mode field expansion for a closed 
cavity \cite{messiah}. For finite damping, however, 
the operators $a_j$ have mixed creation and annihilation character.

In short, we have established an expansion of the quantum
field $\phi$ (and its conjugate momentum ${\hat \phi}$)
in terms of operators $a^{\dagger}_j$ and  $a_j$, and then
obtained equations of motion and commutation relations
for the latter.  This, in principle, completes the program
of second-quantization, and it remains to use these results
to compute correlation functions and propagators, which
we proceed to do in the following Sections.

However, the deviation of the commutators
(\ref{comm-r2}) and (\ref{comm-r2a}) from
the canonical form prevents the construction of
a Fock space, as is the case for quantum dissipative
systems in general \cite{diss}.

\section{Correlation functions}
\label{correl}

The formalism derived in the last Section
for expanding the quantum field $\phi$ in terms
of the operators $a^{\dagger}_j$ and $a_j$ will 
be applied to the calculation of equilibrium correlation functions,
yielding discrete representations for the cavity correlator 
$F$ in 
the presence of dissipation. Section \ref{corr-gen}
investigates the general case, and 
Section \ref{comp} compares the results with those
obtained from the Green's function approach in Section \ref{1comp}.
Section \ref{dos} evaluates and discusses the density of states.

\subsection{General case}
\label{corr-gen}

In equilibrium, the initial conditions for (\ref{eqom-a_n}) 
are irrelevant and the dynamics are completely specified by the 
driving force $b$, i.e.,

\begin{equation}
  a_j(t)=\frac{if_j(a^+)}{2\omega_j}
    \int_{-\infty}^t\!\mskip-0.5\thinmuskip dt'\,e^{i\omega_j(t'-t)}b(t')\;.
\label{ajb}
\end{equation}

\noindent
The nonzero imaginary part of the $\omega_j$ renders 
the integral rapidly converging, in contrast to the conservative case. 
Fourier transforming and taking expectation values then lead to

\begin{equation}
\langle\tilde{a}_j(\omega)\mskip0.5\thinmuskip a_k\rangle=
\frac{f_j(a^+)\mskip0.5\thinmuskip f_k(a^+)}
 {4\omega_j\omega_k(\omega_j-\omega)(\omega_k+\omega)}\,
 \langle\tilde{b}(\omega)\mskip0.5\thinmuskip b\rangle \;.
\label{a_n-a_m}
\end{equation}

\noindent
Since $b$ is fully specified by the {\em in}\/coming waves 
from the {\em free}\/ string $a<x<\infty$, it
does not ``know" about the cavity $x\le a$, so one can use the 
free infinite-string correlation function to calculate its 
spectral density from the definition (\ref{def-b}) as

\begin{eqnarray}
  \langle\tilde{b}(\omega)\mskip0.5\thinmuskip b\rangle&=&-(\partial_x-i\omega)^2
    {\langle\tilde{\phi}(x,\omega)\mskip0.5\thinmuskip\phi(y)\rangle}_{\rm free}
    \bigr|_{y=x}\nonumber\\
  &=&-(\partial_x-i\omega)^2\left.
    \frac{\cos[\omega(x-y)]}{\omega(1-e^{-\beta\omega})}\right|_{y=x}\nonumber\\
  &=&\frac{2\omega}{1-e^{-\beta\omega}}\;.
\label{b-corr}
\end{eqnarray}

For a simple check, antisymmetrize (\ref{a_n-a_m}) 
in $j$ and $k$ and perform the 
inverse Fourier transform to reproduce (\ref{comm-r2}) 
(for the expectation value of the commutator).
Incidentally, by assuming other forms for
$  \langle\tilde{b}(\omega)\mskip0.5\thinmuskip b\rangle $,
the theory accommodates various incoming pump fields.

Given the two-point function (\ref{b-corr}) for the driving force,
it is straightforward to compute the two-point function
for the response, namely the field--field correlation function inside the cavity.
This now merely requires summation, that is, combination of 
(\ref{evolve}), (\ref{a_n-a_m}) and (\ref{b-corr}) leads to

\begin{equation}
  \tilde{F}(x,y,\omega)=\frac{\omega}{1-e^{-\beta\omega}}\sum_{jk}
    \frac{f_j(a^+)\mskip0.5\thinmuskip f_k(a^+)}
         {2\omega_j\omega_k(\omega_j-\omega)(\omega_k+\omega)}
  f_j(x)\mskip0.5\thinmuskip f_k(y)\;.
\label{ff-corr}
\end{equation}

\noindent
The above derivation leads to a clear 
physical interpretation of the pole structure of (\ref{ff-corr}) 
in the complex $\omega$-plane: the Matsubara poles at 
$\omega= i\mu_m = 2i\pi mT$ ($m\in{\bf Z}$) arise  from
the thermal character of the incoming noise, while the QNM 
poles correspond to cavity resonances excited by this noise.

\subsection{Comparison of two forms for the correlation function}
\label{comp}

It will be noticed that we have obtained two
different QNM expansions for ${\tilde F}$, namely 
the double sum in (\ref{ff-corr}) 
and the single sum in (\ref{F-d}).  We next prove their equivalence
without invoking the QNM expansion of a quantum field.

To do so, we rely on the identity \cite{fn7} 

\begin{equation}
  \tilde{G}^{\rm R}(x,y,\omega)-\tilde{G}^{\rm R}(x,y,-\omega)=
  \frac{2\omega}{i}\,\tilde{G}^{\rm R}(x,a^+,\omega)\mskip0.5\thinmuskip
  \tilde{G}^{\rm R}(y,a^+,-\omega)
\label{G-ident}
\end{equation}

\noindent
for $x,y\le a$.  This identity, proved in Appendix \ref{GR-app},
has no nontrivial counterpart in closed, 
conservative systems.  For an interpretation, notice that 
$\tilde{G}^{\rm R}(x,y,\omega)-\tilde{G}^{\rm R}(x,y,-\omega)
\propto\mathop{\rm Im}\nolimits\tilde{G}^{\rm R}(x,y,\omega)$ vanishes in the 
conservative limit and hence is a measure of dissipation, 
which the right hand side states as taking place exclusively at the 
surface $x=a^+$.

Given this identity, the equivalence of the two
expressions for ${\tilde F}$ follows simply by canceling the 
Bose prefactors in (\ref{F-d}) and (\ref{ff-corr}) 
and comparing the result with the Fourier transform of 
(\ref{greenfie}).

Although the two forms are equivalent, each has its
own attractive properties.  The diagonal form
(\ref{F-d}) is simpler, while 
the non-diagonal form  (\ref{ff-corr}) 
is manifestly factorizable: ${\tilde F}(x,y,\omega) 
= A(\omega) \chi(x,\omega) \chi(y,-\omega)$ \cite{fn12}.
Anticipating a similar structure for Feynman propagators,
the non-diagonal form permits  
a quantum in one mode $j$ 
to propagate to another mode $k$,
while the diagonal form implies that the mode index
is ``conserved".  

The expansion of correlations involving $\phi$ alone 
is not unique,
on account of the doubling of QNMs ($j$ and $-j$) compared to NMs
\cite{LTY1,LTY2}.
As discussed in Section \ref{class}, it is more natural 
to consider $\bbox{\phi} = (\phi, {\hat \phi})^{\rm T}$, which leads to a unique
expansion.  Thus we define a tensor field-field
correlator

\def\s{\rule{0mm}{12mm}}
\begin{eqnarray}
  \tilde{\sf F}(x,y,\omega)&\equiv&
   \langle\tilde{\bbox{\phi}}(x,\omega)\mskip0.5\thinmuskip
   \otimes  \mskip0.5\thinmuskip \bbox{\phi}(y)\rangle
\nonumber\\
&=& \s \pmatrix{ 
\langle \phi(x,\omega) \phi(y) \rangle  & 
\langle \phi(x,\omega) {\hat \phi}(y) \rangle \cr
\langle {\hat \phi}(x,\omega) \phi(y) \rangle  & 
\langle {\hat \phi}(x,\omega) {\hat \phi}(y) \rangle }
\nonumber \\
&=& \s \pmatrix{ 1 & i\omega\rho(y) \cr 
                -i\omega\rho(x) & \omega^2\rho(x)\mskip0.5\thinmuskip\rho(y) }
     \tilde{F}(x,y,\omega)\;,
\label{Ftensor}
\end{eqnarray}

\noindent
which can be expressed as 

\begin{equation}
  \tilde{\sf F}(x,y,\omega)=
   \sum_{jk}\tilde{a}_{jk}(\omega)\bbox{f}_{\mskip-0.5\thinmuskip\!j}(x)
   \mskip0.5\thinmuskip \otimes \mskip0.5\thinmuskip
   \bbox{f}_{\mskip-0.5\thinmuskip\!k}(y)\;,
\label{sfF-expand}
\end{equation}

\noindent
where $\tilde{a}_{jk}$ is evaluated to be (Appendix \ref{app-tens})

\begin{equation}
  \tilde{a}_{jk}(\omega)=\frac{\omega f_j(a^+)\mskip0.5\thinmuskip f_k(a^+)}
   {2(1-e^{-\beta\omega})\omega_j\omega_k(\omega_j-\omega)(\omega_k+\omega)}\;;
\label{ajk-res}
\end{equation}

\noindent
that is, the non-diagonal expansion (\ref{ff-corr}) is 
the unique one which generalizes to the tensor~$\tilde{\sf F}$ 
as in (\ref{sfF-expand}).

\subsection{Density of states}
\label{dos}

Another important quantity is the density of states,
which figures prominently in the heuristic argument of
Purcell \cite{purcell} and others \cite{ching}.
The local density of states $d(x,\omega)$,
given below only for real positive $\omega$,
is related to the correlation function $\tilde{F}$ by

\begin{equation}
  d(x,\omega) = -\frac{2\omega}{\pi} \mathop{\rm Im}\nolimits
  {\tilde G}^{\rm R}(x,x,\omega)
= \frac{\omega}{\pi}(1-e^{-\beta\omega})
          \tilde{F}(x,x,\omega) \;,
\label{lDOS}
\end{equation}

\noindent
which allows expression of this important quantity 
in terms of the QNMs. From (\ref{F-d}) one gets

\begin{equation}
  d(x,\omega) = \frac{\omega}{\pi}
  \sum_j\mathop{\rm Im}\nolimits\frac{f_j^2(x)}{\omega_j(\omega_j-\omega)}\;,
\label{lDOS-d}
\end{equation}

\noindent
while the non-diagonal expression
(\ref{ff-corr}) gives

\begin{eqnarray}
  d(x,\omega)&=&\frac{\omega^2}{2\pi}
  \sum_{jk}\frac{f_j(a^+)f_k(a^+)}
{\omega_j\omega_k (\omega_j-\omega)(\omega_k+\omega)}\mskip0.5\thinmuskip
f_j(x) f_k(x)
\label{2DOS-d}\\
&=&\frac{1}{2\pi}
  \sum_{jk}\frac{f_j(a^+)f_k(a^+)}
{(\omega-\omega_j)(\omega+\omega_k)}\mskip0.5\thinmuskip
f_j(x) f_k(x) \;.
\label{2DOS-d2}
\end{eqnarray}

\noindent
The second form results from the first by use of the identity

\begin{equation}
  \sum_j\frac{f_j(x)\mskip0.5\thinmuskip f_j(y)}{\omega_j}=0\;,
\label{QNM-ident}
\end{equation}

\noindent
which follows from (\ref{greenfie}) by letting 
$t\downarrow0$. 

Superficially, the diagonal form is simpler.
However, if we take a single resonance approximation, (\ref{lDOS-d})
yields, with one term $j$

\begin{equation}
  d(x,\omega) \approx \mskip0.5\thinmuskip\frac{\omega}{\pi}
  \mathop{\rm Im}\nolimits\frac{f_j^2(x)}{\omega_j(\omega_j-\omega)}\;,
\label{lDOS-dr}
\end{equation}

\noindent
which is not positive definite.
On the other hand, for the non-diagonal form, the
appropriate approximation is to take one $j$ and $k=-j$
in (\ref{2DOS-d2}), leading to

\begin{equation}
d(x,\omega) \approx \frac{ | f_j(a^+) f_j(x) |^2 }
{2\pi \left[ (\omega- \mathop{\rm Re}\nolimits \omega_j)^2
 + (\mathop{\rm Im}\nolimits \omega_j)^2 \right]} \;,
\label{2DOS-dr}
\end{equation}

\noindent
which is manifestly positive and moreover lorentzian. From this
expression one finds, to leading order in
$\left| \mathop{\rm Im}\nolimits \omega_j \right| = \gamma$,
that

\begin{equation}
\int_{\rm res}\!\!d\omega\int_0^{a^+}\!\!dx\,\rho(x)
\mskip0.5\thinmuskip d(x,\omega)\approx1\;,
\label{unitweight}
\end{equation}

\noindent
where the $\omega$-integral is over one resonance.
This statement is readily derived from (\ref{2DOS-dr}) by
using (\ref{ratiolim}) and the
fact that $\int\!dx\,\rho(x) |f_j(x)|^2  \approx 1$ for a narrow resonance.
Recall that in the modes of the universe approach \cite{ching},
the unit weight of the resonances emerges simply as a numerical result,
and is difficult to understand theoretically.
Here the same result (in 1 d) is justified analytically,
and moreover, one can in principle
(a) estimate the corrections due to other resonances
(note that there is no ``background"
apart from the QNM contributions),
(b) calculate the corrections to higher order in $\gamma$, and
(c) discuss the local density of states $d(x,\omega)$ rather than the
integrated $\int\!dx \,d(x,\omega)$.
Incidentally, this discussion shows that of the two equivalent forms
(\ref{2DOS-d}) and (\ref{2DOS-d2}), the latter is the more appropriate,
since it leads to a finite integral over $\omega$ in the single-resonance
approximation.

One can derive another sum rule,

\begin{equation}
\int_0^{\omega} d(x,\omega')\, d\omega' \approx \frac{\omega}
{\pi \sqrt{\rho(x)}}
\label{sum2}
\end{equation}

\noindent
for large $\omega$.  This second sum rule \cite{ching} states that the
states are merely redistributed without changing their total number.
However, this sum rule is not immediately useful when expressed in terms of
the QNMs, and will not be further discussed here.

\section{Feynman propagator}
\label{feyn}

\subsection{Derivation of the Feynman propagator}

Another important correlation function is the Feynman 
propagator 

\begin{equation}
  G^{\rm F}(x,y,t)=-i\langle{\rm T}\{\phi(x,t)
  \mskip0.5\thinmuskip\phi(y)\}\rangle\;,
  \label{def-Fprop}
\end{equation}

\noindent
in which T denotes time-ordering.
Taking the Fourier transform of the definition 
(\ref{def-Fprop}) leads to a direct relation 
to the correlator (\ref{def-F}),

\begin{equation}
  \tilde{G}^{\rm F}(x,y,\omega)=-\int\!\frac{d\omega'}{2\pi}\,
  \left\{\frac{1}{\omega'+\omega-i\varepsilon}
  +\frac{1}{\omega'-\omega-i\varepsilon}\right\}
  \tilde{F}(x,y,\omega')\;.
\label{GF-F}
\end{equation}

\noindent
We shall limit the discussion below to $T=0$.
Substitution of the right hand side of (\ref{ff-corr}) 
into (\ref{GF-F}) yields $\tilde{G}^{\rm F}$ 
as \cite{fn10} 

\begin{equation}
  \tilde{G}^{\rm F}(x,y,\omega) =-i\sum_{jk}
  \frac{f_j(a^+)\mskip0.5\thinmuskip f_k(a^+)}
      {2\omega_j\omega_k(\omega_j+\omega_k)}
  \left\{\theta(\omega)\frac{\omega_j}{\omega_j-\omega}
         +\theta(-\omega)\frac{\omega_k}{\omega_k+\omega}\right\}
  f_j(x)\mskip0.5\thinmuskip f_k(y)\;.
\label{GF-od}
\end{equation}

\noindent
The cavity Feynman propagator can also be expressed in 
diagonal form, by substituting (\ref{F-d}) into (\ref{GF-F}). 
Again taking $T=0$, this leads to

\begin{eqnarray}
  \tilde{G}^{\rm F}(x,y,\omega)&=&\frac{1}{2}\sum_j
    \frac{f_j(x)\mskip0.5\thinmuskip f_j(y)}{\omega_j(|\omega|-\omega_j)}
\label{GF-d}\\
  &=&\frac{1}{2}\sum_j\frac{f_j(x)\mskip0.5\thinmuskip f_j(y)}
                          {|\omega|(|\omega|-\omega_j)}\;,
\label{GF-d2}
\end{eqnarray}

\noindent
for real $\omega$.  It is stressed that these forms as single sums exist 
even though $\langle T\{a_j(t)\mskip0.5\thinmuskip a_k\}\rangle\not
\propto\delta_{j,\pm k}$ in general. The form (\ref{GF-d2}) 
for $\tilde{G}^{\rm F}$ has been derived from (\ref{GF-d}) 
by means of the QNM identity (\ref{QNM-ident}).
The second form with its divergence \cite{fn11}  
at $\omega=0$ is less convenient than the first.  It has 
been included to show that caution is needed when speaking about 
``the contribution of one QNM". In fact the two summands are almost 
equal if $|\omega|\approx\omega_j$; such resonances are seen 
to be exclusively associated to terms with $j\ge0$.

All of these equivalent expressions (\ref{GF-od}), (\ref{GF-d}) and (\ref{GF-d2})
can be written
generally as

\begin{equation}
\tilde{G}^{\rm F}(x,y,\omega) = \sum_{jk} f_j(x) \Delta_{jk}(\omega) f_k(y) \;,
\label{del1}
\end{equation}

\noindent
with different forms for $\Delta_{jk}$.
This has an obvious diagrammatic interpretation:
the field at $x$ ($y$) couples to the QNM $j$ ($k$)
with a vertex $f_j(x)$ ($f_k(y)$), and the QNM
propagates from mode $j$ to mode $k$ with an amplitude
$\Delta_{jk}$.
This may be compared with the more familiar case
of an infinite conservative system, say

\begin{equation}
\tilde{G}^{\rm F}(x,y,\omega)
= \int\!\frac{dp}{2\pi}\,e^{-ipx}
\Delta(p,\omega)\mskip0.5\thinmuskip e^{ipy} \;.
\end{equation}

\noindent
It is seen that $\int\! dp\, \cdots$ is replaced by $\sum_{jk} \cdots\;$.
An important goal of the present second-quantized 
theory is to study cavity--atom interactions \cite{partII}, 
often referred to as CQED. The objective is to 
establish a set of ``QNM Feynman rules", in which each 
line in a diagram is represented not by a continuous momentum, 
but by one discrete index (or a pair of them) --- not only for 
computational convenience, but also because each term 
can be associated with a cavity 
resonance. Such a discrete representation is especially useful 
for microscopic cavities, where the resonances are widely spaced 
in frequency. The above results are crucial for establishing these 
Feynman rules.

The possibility of alternate expressions for the propagator
may recall a similar situation with gauge theories \cite{bwlee},
though the reasons are quite different.

\subsection{Decay rate and the resonance approximation}
\label{corr-ra}

While the use of the Feynman propagators in an
interacting theory will be presented elsewhere \cite{partII},
it is neverthess profitable at this point to
consider the very simple example of
an atom coupled to the field at a fixed point 
$x$; in the dipole approximation, the decay
rate is related to
the equal-space propagator

\begin{equation}
\tilde{D}(\omega) \equiv \tilde{G}^{\rm F}(x,x,\omega)\;.
\label{eqtprop}
\end{equation}

In particular, we shall be interested in the single-resonance 
approximation (RA) for~$\tilde{D}$.
The obvious choice is to take a single term of the 
sum in (\ref{GF-d}), i.e.,

\begin{equation}
  \tilde{D}(\omega) \approx \tilde{D}_{{\rm ra}'}(\omega)\equiv
    \frac{f_j(x_0)^2}{2\omega_j(|\omega|-\omega_j)}\;.
\label{GFd-ra}
\end{equation}

\noindent
The alternative is to start from (\ref{GF-od}), 
and retain only the $(j,-j)+(-j,j)$ terms 
(only when $k=-j$ does the factor $\omega_j+\omega_k$ in 
the denominator of (\ref{GF-od}) get small close to 
the conservative limit, which is the only case in which 
a single resonance can dominate) to arrive at

\begin{equation}
  \tilde{D}(\omega)\approx\tilde{D}_{\rm ra}(\omega)\equiv
  \frac{|f_j(x_0)\mskip0.5\thinmuskip f_j(a^+)|^2}
  {4|\omega_j|^2\left|\mathop{\rm Im}\nolimits\omega_j\right|}
  \left[\frac{\omega_j}{|\omega|-\omega_j}
        -\frac{\omega_j^*}{|\omega|+\omega_j^*}\right]\;.
\label{Dappr}
\end{equation}

\noindent
Without loss of generality choosing $j>0$, 
the second (non-resonant) term in (\ref{Dappr}) 
is of the same order as those already neglected, and 
hence for most purposes may be omitted.  However, only 
the sum of the two terms in (\ref{Dappr}) preserves the 
fundamental relation\cite{AGD}

\begin{equation}
  \tilde{D}^{\rm R}(\omega)=\tilde{D}^{{\rm A}*}(\omega)
\label{DR-DA}
\end{equation}

\noindent
for real $\omega$, where $\tilde{D}^{\rm R}$ ($\tilde{D}^{\rm A}$) 
is the retarded (advanced) propagator obtained from 
$\tilde{D}(\omega)$ by continuation from positive 
(negative) frequencies. 
As a consequence, it turns out\cite{partII} that
keeping both terms and using the ensuing cavity propagator 
to compute the self-energy of a two-level atom
leads to a renormalization of the level splitting that is guaranteed to be real. 
Of course, for $\tilde{D}_{{\rm ra}'}$ the equality (\ref{DR-DA}) 
is always violated.

Moreover, $\tilde{D}_{{\rm ra}'}$ does not obey the 
equally fundamental inequality  $\mathop{\rm Im}\nolimits\tilde{D}(\omega)\le0$ 
on the real axis \cite{AGD} (see also (\ref{relatedD}) below), 
which $\tilde{D}_{\rm ra}$ satisfies term by term. 
Violation of this inequality in general leads to a retarded atom 
propagator that has poles in the upper half $\omega$-plane \cite{partII}, 
signifying an unphysical instability.

To be sure, in spite of these crucial differences between 
$\tilde{D}_{\rm ra}$ and $\tilde{D}_{{\rm ra}'}$ 
their residues at $|\omega|=\omega_j$ agree in 
the conservative limit, in which the domination of a 
single QNM becomes rigourous. For a proof it suffices 
to note that $|f_j(a^+)|^2/2\left|\mathop{\rm Im}\nolimits
\omega_j\right|\rightarrow1$ 
in this limit (Appendix~\ref{ident-app}). 

We have discussed the single-resonance approximation to
both the density of states $d(x,\omega)$ and to the equal
space propagator ${\tilde D}(\omega)$.  In fact, the arguments are
equivalent, which can be appreciated physically from the fact
that they both relate to the decay rate, and mathematically from
the following identity for real positive $\omega$:

\begin{equation}
d(x,\omega) = -\frac{2\omega}{\pi} \mskip0.5\thinmuskip
 \mathop{\rm Im}\nolimits {\tilde D}(\omega)\;.
\label{relatedD}
\end{equation}

In several places we have remarked that the non-diagonal
QNM representation has some nice properties,
and is in fact the unique representation if the field $\phi$ and the
conjugate momentum ${\hat \phi}$ are considered together,
for example in the tensor correlator (\ref{sfF-expand}).
There are of course many ways to understand why the correlator
is non-diagonal; one of the most direct is via (\ref{eqom-a_n}),
which shows that all the mode coefficients $a_j$ are driven by
the same force $b(t)$, so in general different coefficients will
have phase coherence and hence a nonzero correlation.
Incidentally, this non-diagonal nature is {\em not}\/
a quantum effect, since the property survives at high temperatures, e.g.,
$\beta \rightarrow 0$ in (\ref{ff-corr}).
However, all the propagators and correlation functions
become diagonal in the conservative limit, as they should.
In fact, applying (\ref{ortho}) to (\ref{a_n-a_m}) in this 
limit readily yields

\begin{equation}
  \langle\tilde{a}_j(\omega)\mskip0.5\thinmuskip a_k\rangle=
  \frac{\pi}{\omega(1-e^{-\beta\omega})}
  \mskip0.5\thinmuskip\delta(\omega-\omega_j)\mskip0.5\thinmuskip
  \delta_{j,-k}\;,\label{cons-corr}
\end{equation}

\noindent
in agreement with the creation--annihilation interpretation of the $a_j$
in this limit given above (\ref{nm-exp}). As a result both
(\ref{ff-corr}) for $\tilde{F}$ and (\ref{GF-od}) for
$\tilde{G}^{\rm F}$ become diagonal in the conservative limit
as well.

\section{Example: the dielectric rod}
\label{example} 

A useful check and example of the preceding is given by the 
``dielectric rod" model \cite{lly}:

\begin{equation}
  \rho(x)=n^2\mskip0.5\thinmuskip\theta(a-x)
  +n_0^2\mskip0.5\thinmuskip\theta(x-a)\;.
\label{rho-dr}
\end{equation}

\noindent
That is, we generalize the condition (\ref{rho-out}) 
and allow $\rho(x{>}a)$ to be an arbitrary constant $n_0^2$. 
To be sure, this generalization is trivial in principle since a 
model with parameters $(n,n_0,a)$ can be mapped onto one with 
parameters $(n/n_0,1,n_0a)$ by the substitution $x\mapsto n_0x$. 
Yet it is convenient in practice, since we can now deal with
two different conservative limits 
(see the discussion below (\ref{dr-f})): $n/n_0\rightarrow0$
(the ``nodal" limit) and 
$n/n_0\rightarrow\infty$ (the ``antinodal" limit), by letting 
$n_0\rightarrow\infty$ and $n_0\rightarrow0$, respectively, 
while keeping $n$, $a$, and hence $\mathop{\rm Re}\nolimits\omega_j$ 
(see (\ref{dr-omega}) below) fixed.

The model (\ref{rho-dr}) can be solved exactly for the 
QNM frequencies \cite{LTY1}, which read \cite{fn8}

\begin{eqnarray}
  na\omega_j&=&\Biggl\{  \begin{array}{ll}
  (j+{\textstyle\frac{1}{2}})\pi-i\mathop{\rm acth}\nolimits(n/n_0),&n>n_0\;;\\
  j\pi-i\mathop{\rm ath}\nolimits(n/n_0),&n<n_0 \end{array}
\label{dr-omega} \\
  &=&j\pi-\frac{i}{2}\ln\frac{n_0+n}{n_0-n}\;.
\end{eqnarray}

\noindent
Both $n_0\rightarrow\infty$ and $n_0\rightarrow0$ are 
indeed seen to be conservative limits ($\mathop{\rm Im}\nolimits
\omega_j \rightarrow 0$).
They correspond to 
clamped and free ends, respectively, in the  
interpretation of the wave equation as the transverse vibrations of
a string \cite{dekker,laistring}. On the other hand, 
for $n_0 \rightarrow n$ the QNM description breaks down as 
the dissipation tends to infinity.

The QNM wavefunctions are given inside the cavity by \cite{fn8}

\begin{equation}
  f_j(x)=\sqrt{\frac{2}{n^2a}}\sin(n\omega_jx)\;;
\label{dr-f}
\end{equation}

\noindent
their normalization is still given by (\ref{normal}), 
but our generalization $\rho(x{>}a)=n_0^2$ is readily shown 
to imply a corresponding modification of the surface term in 
the scalar product definition itself, viz.,

\begin{equation}
  \langle\bbox{\phi},\bbox{\chi}\rangle=i \left\{
  \int_0^{a^+}\!\!dx\,\left[\phi(x)\mskip0.5\thinmuskip
  \hat{\chi}(x)+\hat{\phi}(x)\mskip0.5\thinmuskip\chi(x)\right]
  +n_0\mskip0.5\thinmuskip\phi(a^+)\mskip0.5\thinmuskip\chi(a^+)\right\}\;.
\label{scalarpr2}
\end{equation}

In the ``nodal" limit $n_0\rightarrow\infty$, (\ref{dr-omega}) 
and (\ref{dr-f}) show that $f_j(a)\sim n_0^{-1}$, so that 
the factor $f_j(a)\mskip0.5\thinmuskip f_k(a)$ in the surface term of the 
orthogonality relation (\ref{ortho}), (\ref{scalarpr2}) 
overcomes the explicit factor $n_0$, allowing the surface 
term to be neglected. This nodal limit has a counterpart in 
the ``loaded string" model $\rho(x)=1+M\delta(x-a)$, where $M$ 
can be set to infinity \cite{fn9b}. On the other hand, in the antinodal 
limit $n_0\rightarrow0$ it is the explicit $n_0$ which allows 
neglect of the surface term, $f_j(a)$ tending to a constant. 
Hence, the QNM expansion becomes a standard normal-mode expansion 
also if $n_0\ll n$, which clarifies a detail left open in 
Section \ref{class} and Ref. \cite{LTY1}.

By means of a partial fraction expansion and the 
identity (\ref{QNM-ident}), the expression (\ref{F-d}) 
for $\tilde{F}$ can be rewritten as

\begin{equation}
  \tilde{F}(x,y,\omega)=\frac{-i}{2\omega(1-e^{-\beta\omega})}
  \sum_jf_j(x)\mskip0.5\thinmuskip f_j(y)
  \left[\frac{1}{\omega_j-\omega}+\frac{1}{\omega_j+\omega}\right]\;,
  \label{F-d2}
\end{equation}

\noindent
which is analytically more convenient even though the 
sum over $j$ converges more slowly. Upon substitution of 
the dielectric rod QNMs (\ref{dr-omega}) and (\ref{dr-f}) 
into (\ref{F-d2}), the sum over $j$ can be performed as 
the {\em conventional}\/ Fourier series \cite{fn9a}

\begin{equation}
  \sum_j\frac{e^{-ji\pi z}}{j\pi-i\alpha}=
  \frac{2ie^{\alpha z}}{e^{2\alpha}-1}\;,\qquad0<z<2\\
\end{equation}

\noindent
implying

\begin{equation}
  \sum_j\frac{e^{ji\pi z}}{j\pi-i\alpha}=
  \frac{2ie^{\alpha(2-z)}}{e^{2\alpha}-1}\;,\qquad0<z<2\;,
\label{fouriersum}
\end{equation}

\noindent
and some rearrangement yields the correlation function as

\begin{equation}
  \tilde{F}(x,y,\omega)=\frac{2n_0\sin(n\omega x)\sin(n\omega y)}
  {\omega(1-e^{-\beta\omega})
   [n_0^2\sin^2(n\omega a)+n^2\cos^2(n\omega a)]}\;.
\label{F-dr}
\end{equation}

\noindent
This result also follows from the 
modes of the universe approach
in Appendix \ref{MU-app}.

As discussed in Section \ref{1comp}, the subtracted
correlation function $F_{\rm S}$ is directly related to the energy density,
and is in fact the squared amplitude of the field strength.
Figure~\ref{fig1} shows $F_{\rm S}(x,x,t)$ versus $x$ at $t = 0.1$
for the dielectric rod model
with $a=1, n_0=1, n=5$, for different values of $\beta$; this shows
that the field amplitude is largest near the leaky end of the rod.
Figure~\ref{fig2} shows $F_{\rm S}(x,x,t)$ versus $t$ at $x = 0.3$ (all other
parameters the same as before).  This diagram vividly illustrates the
advantage of the QNM approach --- although the result is in principle
obtainable from the modes of the universe method,
the clear oscillatory signal is best
captured by expressing this quantity in terms of QNMs.  

Also, for the Feynman propagator the sum (\ref{GF-d}) can be 
performed if $\rho={\rm const}$, yielding

\begin{equation}
  \tilde{G}^{\rm F}(x,y,\omega)=
  -\frac{\sin(n\omega x)}{n\omega}
  \frac{n\cos[n\omega(a-y)]-in_0\sin[n|\omega|(a-y)]}
       {n\cos(n\omega a)-in_0\sin(n|\omega|a)}
\label{GF-dr}
\end{equation}

\noindent
for $x<y$, while for $x>y$ the propagator is obtained 
via $\tilde{G}^{\rm F}(x,y,\omega)=\tilde{G}^{\rm F}(y,x,\omega)$. 

Notice that the final expressions (\ref{F-dr}) and (\ref{GF-dr}) tend to a
finite limit if $n_0\rightarrow n$ even though the individual terms in
 (\ref{F-d}) and (\ref{GF-d}) do not.  In this semi-infinite string
limit the very notions of cavity and environment lose their meaning,
and indeed the right hand sides of (\ref{F-dr}) and (\ref{GF-dr}) are seen
to become independent of $a$.

\section{Final remarks}
\label{remarks}

To summarize, we have developed the second-quantized version
of the field theory using the QNM basis.  Various physical
quantities are then written as sums over QNM contributions
--- either as diagonal sums over a single index $j$, or as non-diagonal
sums over a pair of indices $jk$.  The resonance approximation
is studied, leading to a proof of the unit weight of narrow
resonances in the density of states, or equivalently the enhancement
rate for the decay of excited states as embodied in the behavior
of the equal space propagator ${\tilde D}(\omega)$.

As has been mentioned already in Section \ref{intro}, an
important extension of the present work is to include 
matter in the hamiltonian (\ref{def-H}), enabling the 
application of QNMs to quantum optics. This will be the 
subject of the sequel paper Ref. \cite{partII}. Other 
generalizations include the study of vector fields, 
and of open systems in three space dimensions. Further, 
a development parallelling the present one could be carried 
out for the Klein--Gordon equation instead of the wave 
equation (\ref{wave-eq}). Since the two evolution equations 
are directly related by a transformation of the spatial variable \cite{LTY1}
however, this has not been taken up here.

Instead of generalizing the physical system one can 
also relax the assumption of global equilibrium made 
in Section \ref{correl}. It is recalled here that
the formalism of Sections \ref{incoming} and \ref{quant} --- and 
in particular the driving force $b$ of (\ref{def-b}) ---
is well-defined for any initial state of the fields; taking 
a coherent state for the latter instead of a thermal one 
enables the study of a pumped cavity.

On the theoretical side, it would be interesting to 
provide a path-integral formulation of QNM quantization. 
This can supposedly be done on two levels. The first, 
semi-phenomenological one is to write down an effective 
action generating dynamics equivalent to  
(\ref{eqom-a_n}). The second, more fundamental one 
is to start with the action for the whole universe for our model (\ref{def-H}), 
integrate out the degrees of freedom of the outside, 
and use a QNM basis for the ensuing dynamics of the cavity.

For such future developments, this article can 
hopefully serve as a starting point and reference. 
In conclusion, we have shown that the QNM expansion 
is as powerful for open second-quantized systems as 
it is for their classical counterparts.

\section*{Acknowledgment}

This work is supported by Grant No.\ 452/95P of 
the Hong Kong Research Grants Council.
We thank C. K. Au, E. S. C. Ching,
W. M. Suen and C. P. Sun for discussions.

\appendix

\section{Relation between surface term and imaginary part of the frequency}
\label{ident-app}

In this Appendix we give an alternative proof of the identity
(\ref{ratiolim}) in the conservative limit. For this purpose,
generalize to complex classical fields and define the energy density
$h(x)=|\partial_x\phi|^2/2+|\hat{\phi}|^2/2\rho(x)$, so that
$\dot{h}(x)=-\partial_x j(x)$ with 
the current $j(x)=-\mathop{\rm Re}\nolimits
\left( \hat{\phi}(x)\mskip0.5\thinmuskip  \partial_x\phi(x) \right)
/\mskip-0.5\thinmuskip\rho(x)$. Define the
cavity energy $E=\int_0^{a^+}\!\!dx\,h(x)$, then for a field
$\phi(x,t)=f_j(x)\mskip0.5\thinmuskip e^{-i\omega_jt}$ at $t=0$ one has
\begin{equation}
  E=\frac{\gamma}{2}|f_j(a^+)|^2
    +(\mathop{\rm Re}\nolimits\omega_j)^2\int_0^{a^+}\!\!dx\,
\rho(x)\mskip0.5\thinmuskip|f_j(x)|^2\;,
\end{equation}
where $\gamma\equiv\left|\mathop{\rm Im}\nolimits\omega_j\right|$.
In the conservative limit the first
term vanishes, while the integral in the second term tends to unity so
that $E\rightarrow|\omega_j|^2$. Combination with $-\dot{E}=2\gamma
E=j(a^+)=|\omega_j|^2|f_j(a^+)|^2$ shows that
$|f_j(a^+)|^2/2\gamma\rightarrow1$ in this limit, which proves our
assertion.

\section{Identity for retarded propagator}
\label{GR-app}

In this Appendix we derive the Green's function identity
(\ref{G-ident}). To this end, define $f(x,\omega)$ ($g(y,\omega)$)
as the solution of the homogeneous wave equation (\ref{def-QNM})
(upon
the substitution $\omega_j\mapsto\omega$) satisfying the first
(second) of the boundary conditions (\ref{f-bc}) \cite{fn21}.
This allows one to write

\begin{equation}
  \tilde{G}^{\rm R}(x{<}y,\omega)=
    \frac{f(x,\omega)\mskip0.5\thinmuskip g(y,\omega)}{W(\omega)}\;,
\end{equation}

\noindent
where one can choose $f(x,\omega)=f(x,-\omega)=f^*(x,\omega)$ and
$g(y,\omega)=g^*(y,-\omega)$, and where $W$ is the position-independent
wronskian of the functions $f$ and $g$\cite{lly}. Then one has

\begin{eqnarray}
  \frac{\tilde{G}^{\rm R}(x,y,\omega)-\tilde{G}^{\rm R}(x,y,-\omega)}
    {\tilde{G}^{\rm R}(x,a^+,\omega)\mskip0.5\thinmuskip
     \tilde{G}^{\rm R}(y,a^+,-\omega)}
  &=&\frac{g(y,\omega)W^*(\omega)-g^*(y,\omega)W(\omega)}
         {|g(a^+,\omega)|^2f(y,\omega)}\nonumber\\
  &=&\frac{g(y,\omega)\mskip0.5\thinmuskip{g'}^*(y,\omega)
      -g'(y,\omega)\mskip0.5\thinmuskip g^*(y,\omega)}
         {|g(a^+,\omega)|^2}\;.
\end{eqnarray}

\noindent
The numerator of this last expression is itself another wronskian, and
hence can be evaluated at $y=a^+$ to yield

\begin{eqnarray}
  \frac{\tilde{G}^{\rm R}(x,y,\omega)-\tilde{G}^{\rm R}(x,y,-\omega)}
    {\tilde{G}^{\rm R}(x,a^+,\omega)\mskip0.5\thinmuskip
     \tilde{G}^{\rm R}(y,a^+,-\omega)}
  &=&-2i\mathop{\rm Im}\nolimits
     \frac{g'(a^+,\omega)}{g(a^+,\omega)}\nonumber\\
  &=&\frac{2\omega}{i}\;,
\end{eqnarray}

\noindent
completing the proof of (\ref{G-ident}).

\section{Expansion of tensor correlator}
\label{app-tens}

The coefficients $\tilde{a}_{jk}$ in (\ref{sfF-expand}) are given
by the projection

\begin{equation}
  \tilde{a}_{jk}(\omega)=
  \frac{\langle\!\langle\tilde{\sf F}(\omega),
        \bbox{f}_{\mskip-0.5\thinmuskip\!j}
     \bbox{f}_{\mskip-0.5\thinmuskip\!k}\rangle\!\rangle}
       {4\omega_j\omega_k}\label{tensor-proj}
\end{equation}

\noindent
in terms of the bilinear form on the product space, which reads

\begin{eqnarray}
  \langle\!\langle{\sf P},{\sf Q}\rangle\!\rangle&\equiv&
  -\int_0^{a^+}\!\!dx\mskip0.5\thinmuskip dy\,\bigl\{
   P_{11}(x,y)\mskip0.5\thinmuskip Q_{22}(x,y)
   +P_{12}(x,y)\mskip0.5\thinmuskip Q_{21}(x,y)\nonumber\\
  &&\hphantom{-\int_0^{a^+}\!\!dx\mskip0.5\thinmuskip dy\,\bigl\{}
    +P_{21}(x,y)\mskip0.5\thinmuskip Q_{12}(x,y)+P_{22}(x,y)\mskip0.5\thinmuskip Q_{11}(x,y)\bigr\}\nonumber\\
  &&-\int_0^{a^+}\!\!dx\,\left\{P_{11}(x,a^+)\mskip0.5\thinmuskip Q_{21}(x,a^+)
    +P_{21}(x,a^+)\mskip0.5\thinmuskip Q_{11}(x,a^+)\right\}\nonumber\\
  &&-\int_0^{a^+}\!\!dy\,\left\{P_{11}(a^+\!,y)
     \mskip0.5\thinmuskip Q_{12}(a^+\!,y)
    +P_{12}(a^+\!,y)\mskip0.5\thinmuskip Q_{11}(a^+\!,y)\right\}\nonumber\\
  &&-P_{11}(a^+\!,a^+)\mskip0.5\thinmuskip Q_{11}(a^+\!,a^+)\;.
\end{eqnarray}

\noindent
Substitution of (\ref{Ftensor}) for $\tilde{\sf F}(\omega)$ into
(\ref{tensor-proj}) yields
\begin{eqnarray}
  \tilde{a}_{jk}(\omega)=\frac{1}{4\omega_j\omega_k}
      \biggl\{&&\!(\omega_j+\omega)(\omega_k-\omega)
      \int_0^{a^+}\!\!dx\mskip0.5\thinmuskip dy\,
    \rho(x)\mskip0.5\thinmuskip\rho(y)\mskip0.5\thinmuskip f_j(x)f_k(y)
      \tilde{F}(x,y,\omega)\nonumber\\
    &&+i(\omega_j+\omega)f_k(a^+)
      \int_0^{a^+}\!\!dx\,\rho(x)\mskip0.5\thinmuskip
    f_j(x)\tilde{F}(x,a^+\!,\omega)\nonumber\\
    &&+i(\omega_k-\omega)f_j(a^+)
      \int_0^{a^+}\!\!dy\,\rho(y)\mskip0.5\thinmuskip
    f_k(y)\tilde{F}(a^+\!,y,\omega)\nonumber\\
    &&-f_j(a^+)f_k(a^+)\tilde{F}(a^+\!,a^+\!,\omega)\biggr\}\;.\label{def-ajk}
\end{eqnarray}

\noindent
Inserting {\em any\/} QNM expansion
$\tilde{F}(x,y,\omega)=\sum_{lm}\tilde{b}_{lm}(\omega)f_l(x)f_m(y)$
such as  (\ref{F-d}) or (\ref{ff-corr}) and invoking the relations
(\ref{normal}), (\ref{ortho}) and (\ref{QNM-ident}),
the expression (\ref{def-ajk}) can be evaluated as in (\ref{ajk-res}), which
is what we set out to show.

\section{Modes of the universe approach to the correlation function}
\label{MU-app}

It is instructive to rederive the correlator $F$ using the modes of the
universe (MU). The MU expansion of the fields reads \cite{fn22}

\begin{equation}
  \left(\begin{array}{c} \phi(x)\\ \hat{\phi}(x) \end{array}\right)=\sum_{l>0}
  \left(\begin{array}{c} (u_l^{\dagger}+u_l^{\vphantom{\dagger}})/\sqrt{2\nu_l} \\
    i\rho\sqrt{\nu_l/2}(u_l^{\dagger}-u_l^{\vphantom{\dagger}})
  \end{array}\right)\psi(x,\nu_l)\;,\label{univ-exp}
\end{equation}

\noindent
i.e., it is of the same form as (\ref{nm-exp}) but in
(\ref{univ-exp}) the sum runs over the MU frequencies
$\nu_l=l\pi/\Lambda$ (to leading order in $a/\Lambda\ll1$), the $u_l$
and $\psi_l$ being MU annihilation operators and wavefunctions,
respectively. Insertion of (\ref{univ-exp}) into (\ref{def-F})
yields

\begin{eqnarray}
  \tilde{F}(x,y,\omega)&=&\sum_{lm}
    \frac{\psi(x,\nu_l)\mskip0.5\thinmuskip\psi(y,\nu_m)}
    {2\sqrt{\omega_l\omega_m}}\mskip0.5\thinmuskip
    \bigl<\{\tilde{u}_l^{\dagger}(\omega)
    +\tilde{u}_l^{\vphantom{\dagger}}(\omega)\}
    \{u_m^{\dagger}+u_m^{\vphantom{\dagger}}\}\bigr>\nonumber\\
  &=&\sum_{l}\frac{\psi(x,\nu_l)\mskip0.5\thinmuskip\psi(y,\nu_l)}
     {2\omega_l}\mskip0.5\thinmuskip
    \langle\tilde{u}_l^{\dagger}(\omega)
    \mskip0.5\thinmuskip u_l^{\vphantom{\dagger}}
           +\tilde{u}_l^{\vphantom{\dagger}}(\omega)
    \mskip0.5\thinmuskip u_l^{\dagger}\rangle\nonumber\\
  &=&\frac{\Lambda}{|\omega|}\psi(x,|\omega|)
    \mskip0.5\thinmuskip\psi(y,|\omega|)
    \{\theta(-\omega)N(-\omega)+\theta(\omega)[N(\omega)+1]\}\nonumber\\
  &=&\frac{\Lambda\psi(x,|\omega|)\mskip0.5\thinmuskip\psi(y,|\omega|)}
         {\omega(1-e^{-\beta\omega})}\;.\label{F-MU}
\end{eqnarray}

\noindent
To arrive at the third line, we used
$\tilde{u}_l^{(\dagger)}(\omega)=2\pi\delta(\omega
\mathbin{\raisebox{-2.5pt}{$\stackrel{\raisebox{-3pt}[0pt][0pt]
{{\scriptsize(}\raisebox{-1pt}{+}{\scriptsize)}}}{-}$}}\nu_l)\mskip0.5\thinmuskip
u_l^{(\dagger)}$
and defined the boson occupation number
$N(\omega)=[\exp(\beta\omega)-1]^{-1}$. Comparison of
(\ref{ff-corr}) and (\ref{F-MU}) elucidates why the former factorizes
with respect to $x$ and $y$: this is seen to be a consequence of the
nondegeneracy of the MU spectrum of the semi-infinite string, as
opposed to, e.g., a free string or one with periodic boundary conditions.
Evaluation of $\psi$ for the ``dielectric rod" of Section \ref{example}
at once shows that (\ref{F-MU}) indeed reduces to (\ref{F-dr})
in this case, providing a further comparison between the various
techniques of this paper.

Finally, in terms of the MU the local density of states is defined as

\begin{eqnarray}
  d(x,\omega)&\equiv&\sum_l|\psi(x,\nu_l)|^2\delta(\nu_l-\omega)\nonumber\\
        &=&\frac{\Lambda}{\pi}\psi(x,\omega)^2\;,\label{lDOS-MU}
\end{eqnarray}

\noindent
and comparison with (\ref{F-MU}) at once reproduces (\ref{lDOS}).

\begin{figure}
  \caption{Equal-space correlation function within the dielectric rod as a
           function of $x$ at $t=0.1$ and different inverse temperatures
           $\beta$. The
           refractive indices are $n_0=1$, $n=5$; the width $a=1$.}
  \label{fig1}
\end{figure}
\begin{figure}
  \caption{Equal-space correlation function within the dielectric rod as a
           function of $t$ at $x=0.3$ and different inverse temperatures
           $\beta$. The
           refractive indices are $n_0=1$, $n=5$; the width $a=1$.}
  \label{fig2}
\end{figure}


\begin{thebibliography}{99}
\bibitem{fn3}
See Ref.\cite{LTY1} for the relation 
between the two classes of dissipative systems:
one in which the coupling is mediated through a term in 
the total hamiltonian, and the other in which
the coupling is mediated by a boundary condition ---
the continuity of $\phi$ across $x=a$ in the present case.
\bibitem{fn1}
The technical 
condition that $\phi$ and ${\hat \phi}/\mskip-0.5\thinmuskip\rho$ be 
differentiable, as well as the speed of 
convergence of the sum in (\ref{evolve}), 
are outside the present scope; see Ref.\cite{LTY1}.
\bibitem{fn4}
The fact that the sum in 
(\ref{evolve}) converges more slowly if 
$\bbox{\phi}\not\in\Gamma$\cite{LTY1} is 
irrelevant in the present formal development. 
In any case, for operator fields the large wavenumber 
components never get small as they do for smooth classical 
fields, irrespective of the expansion method. 
For numerical applications, it is best to optimize 
the final formulas for observable physical quantities 
for convergence along the lines of Section III of Ref. \cite{LTY2}. 
Notice, however, that comparison of methods B and C in Ref. \cite{LTY1} 
shows the surface term in (\ref{def-a_n}) to
improve convergence also for function pairs outside 
$\Gamma$, in particular, to be essential for 
convergence to the second component.
\bibitem{fn5}
However, we shall not notationally distinguish between 
corresponding classical and quantized fields, since the
meaning will be clear from the context.
\bibitem{fn7}
For simplicity of notation we 
discuss (\ref{G-ident}) for real $\omega$ only, but its 
proof can be extended to complex $\omega$ by continuing both 
sides analytically.
\bibitem{fn12}
This expression is actually symmetric under $x \leftrightarrow y$,
as will be shown in Appendix~\ref{MU-app}.
\bibitem{fn10}
The equations (\ref{GF-od}), (\ref{GF-d}), (\ref{GFd-ra}) and (\ref{Dappr})
apply for real $\omega$, and the usual care \cite{AGD} 
is needed when performing an analytic continuation from either 
positive or negative frequencies.
\bibitem{fn11}
Per term; after 
summation over $n$ the factor multiplying $|\omega|^{-1}$ 
is seen to vanish near $\omega=0$, as required by consistency.
\bibitem{fn8}
In the case $n>n_0$ 
our numbering violates the convention $\omega_{-j}=-\omega_j^*$. 
Likewise (\ref{dr-f}) violates the sign convention given earlier.
However, these cause no problems here.
\bibitem{fn22}
In (\ref{nm-exp}), the normalization (\ref{normal}) forces the $f_j$
to be real in the conservative limit, in contrast to the standard
formalism where their phase is arbitrary. For uniformity we choose the
$\psi$ in (\ref{univ-exp}) to be real as well; this shows that the
result in (\ref{F-MU}) is well-defined.
\end{thebibliography}
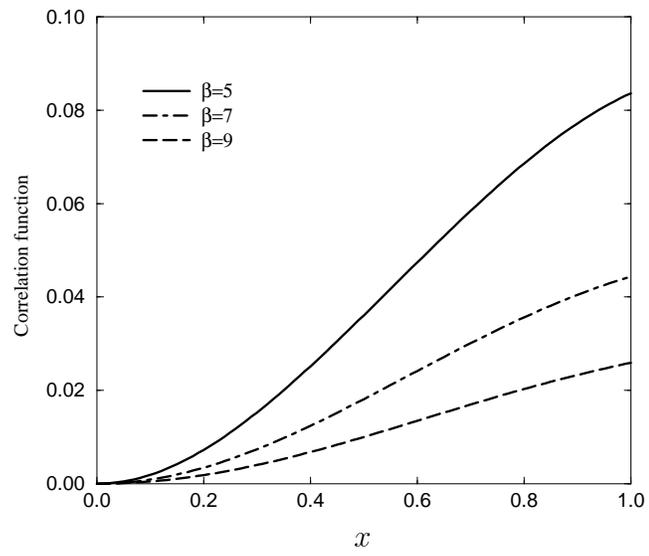
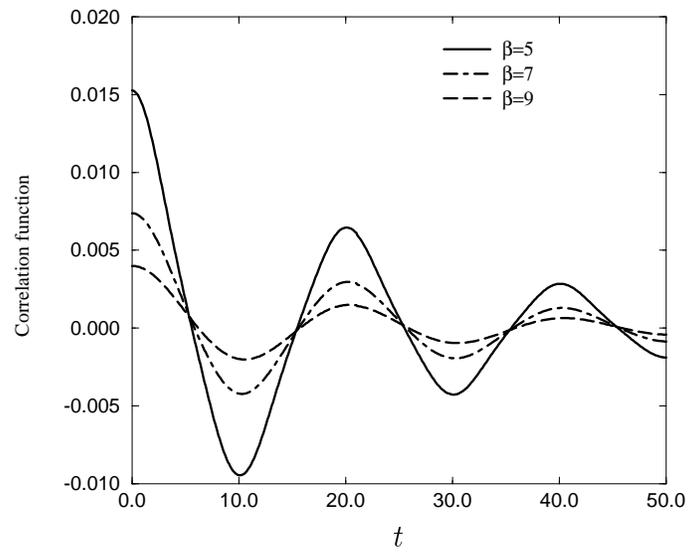
\end{document}